\documentclass[conference]{IEEEtran}
\IEEEoverridecommandlockouts
\usepackage{graphicx}

\usepackage{subcaption}
\usepackage{amsmath,amsthm,amssymb}
\usepackage{floatflt}  
\usepackage[margin=1cm]{caption}
\usepackage[left=1.62cm,right=1.62cm,top=1.7cm]{geometry}

\usepackage{lineno,hyperref}
\usepackage{array}  
\usepackage{setspace}

\usepackage{comment}
 
\usepackage{blindtext}

\usepackage{algorithm}
\usepackage{algpseudocode}  

\usepackage{listings}
\usepackage{xcolor}
\usepackage{scalerel}

\usepackage{algpseudocode}
\usepackage{lipsum}

\def\BibTeX{{\rm B\kern-.05em{\sc i\kern-.025em b}\kern-.08em T\kern-.1667em\lower.7ex\hbox{E}\kern-.125emX}}

\usepackage{flushend}
\IEEEtriggeratref{19}

\begin{document}

\title{EGAN: Evolutional GAN for Ransomware Evasion \thanks{\textcopyright 2023 IEEE. Personal use of this material is permitted. Permission from IEEE must be obtained for all other uses, in any current or future media, including reprinting/republishing this material for advertising or promotional purposes, creating new collective works, for resale or redistribution to servers or lists, or reuse of any copyrighted component of this work in other works.}
}

\author{
\IEEEauthorblockN{1\textsuperscript{st} Daniel Commey}
\IEEEauthorblockA{\textit{Dept. Multidisciplinary Engineering} \\
\textit{Texas A\&M University}\\
Texas, USA\\
dcommey@tamu.edu
}
\and
\IEEEauthorblockN{2\textsuperscript{nd} Benjamin Appiah}
\IEEEauthorblockA{\textit{Dept. Computer Science} \\
\textit{Ho Technical University}\\
Ho, Ghana\\
bappiah@htu.edu.gh
}
\and
\IEEEauthorblockN{3\textsuperscript{rd} Bill K. Frimpong}
\IEEEauthorblockA{\textit{Dept. Computer Science} \\
\textit{Ho Technical University}\\
Ho, Ghana\\
bfrimpong@htu.edu.gh
}
\and
\IEEEauthorblockN{4\textsuperscript{th} Isaac Osei}
\IEEEauthorblockA{\textit{Dept. Computer Science} \\
\textit{Ho Technical University}\\
Ho, Ghana\\
iosei@htu.edu.gh
} 
\and
\IEEEauthorblockN{5\textsuperscript{th} Ebenezer N. A. Hammond}
\IEEEauthorblockA{
\textit{Building \& Road Research Institute}\\
Kumasi, Ghana \\
eahammond@csir.brri.org
}
\and
\IEEEauthorblockN{6\textsuperscript{th} Garth V. Crosby}
\IEEEauthorblockA{\textit{Dept. Engineering Technology \& Industrial Distribution} \\
\textit{Texas A\&M University}\\
Texas, USA\\
gvcrosby@tamu.edu
}
}

\maketitle

\begin{abstract} 
Adversarial Training is a proven defense strategy against adversarial malware. However, generating adversarial malware samples for this type of training presents a challenge because the resulting adversarial malware needs to remain evasive and functional. This work proposes an attack framework, EGAN, to address this limitation. EGAN leverages an Evolution Strategy and Generative Adversarial Network to select a sequence of attack actions that can mutate a Ransomware file while preserving its original functionality. We tested this framework on popular AI-powered commercial antivirus systems listed on VirusTotal and demonstrated that our framework is capable of bypassing the majority of these systems. Moreover, we evaluated whether the EGAN attack framework can evade other commercial non-AI antivirus solutions. Our results indicate that the adversarial ransomware generated can increase the probability of evading some of them.
\end{abstract}


\section{Introduction}
\label{intr}
\par Recent research \cite{Quertier,Song,DemetrioCBLAR21,Demetrio,Li2020ArmsRI,Abdullah,Wei,LingweiChen,Weiwei} has demonstrated that current Machine Learning (ML) or Deep Learning-based malware detection models are inherently vulnerable to adversarial attacks. These attacks typically take the form of adversarial instances, which are intentionally constructed by altering actual inputs. 
\par A robust defense against adversarial malware can be built if the training data is sourced from a variety of inputs, meaning the training data includes samples of this adversarial malware. This type of training is referred to as Adversarial Training \cite{Goodfellow,Szegedy}. However, the strength of Adversarial Training lies in the production of feature-rich training data. In this paper, adversarial malware instances that have evaded the majority of multi-engine scanners and malware sandboxes are identified as such feature-rich data. Nonetheless, due to the complexity of software files, such as the structure of Windows portable executable (PE) files, finding effective ways to create or alter malware instances into their adversarial states for Neural Network training without affecting their functionality has proven to be a challenge \cite{Song,Wei,Abdullah,LingweiChen,Weiwei}.
\par This paper introduces EGAN, an attack system that integrates an Evolution Strategy (ES) learning agent and a Generative Adversarial Network to produce adversarial Ransomware samples. In this system, an ES agent confronts a Ransomware classifier and decides on a series of functionality-preserving actions to apply to Ransomware samples. The approach identifies the most optimal sequence of actions that leads to misclassification for each given Ransomware sample. If the ES agent's manipulations prove ineffective, a GAN is used to generate an adversarial feature vector that alters the Ransomware file to appear benign.
\par According to our experimental results on standard Ransomware samples, the Ransomware generated successfully evaded several static commercial AI-powered anti-virus solutions on VirusTotal. We tested the attack's capacity to bypass various commercial antivirus detectors that use static engines. The test results show that adversarial Ransomware, created using EGAN, can maintain its functionality and evade the majority of static and dynamic detectors.
\par The rest of this work is structured as follows: Section~\ref{rela} introduces the associated context of the proposed work. Section~\ref{REGAN} describes the adversarial Ransomware generation framework. Section~\ref{exp} discusses the data collected, the experimental setup, model implementation, and results. Section~\ref{concl} contains the discussion and conclusions. 

\section{Related Background}
\label{rela} 

\subsection{Adversarial Ransomware samples}
\label{pd}
\par Considering a classification task with input $ x $ and class label $ y $, we identify a perturbation $ \delta $ on input $ x $ such that $ \arg\max_{i \in y} f_{i}(x)\neq y $. The adversarial Ransomware attack aims to optimize the following objective:
\begin{equation}
\label{max}
\max_{\delta} \mathcal{L}(f(x+\delta),y).
\end{equation}
\par Here $ \mathcal{L} $ represents a loss function (typically the cross-entropy), and $ f $ is the classification function. Given access to the gradient of the network $ f $, the attacker targets a label $ y_{i} $ by maximizing $ - \mathcal{L}(f(x+\delta),y_{i}) $. In other words, they seek the best parameter $ \delta $ that will lead to misclassification.
\par Some subsequent studies \cite{Quertier,Demetrio,Li2020ArmsRI,Song,Abdullah,DemetrioCBLAR21,Wei,LingweiChen,Weiwei} have demonstrated that an attacker can work with a black-box learning model to compute the samples without knowing the gradient of $ f $. To be more precise, the attacker can emulate a model using the estimated boundary by predicting the border of the decision region of the model based on the variation in model output triggered by different samples. Subsequently, the parameters of this substitute model are used to generate the adversarial Ransomware.

\subsection{Evolution Strategy}
\label{es}
\par Evolution Strategies (ES) is a type of black-box optimization technique that is inspired by natural evolution: A population of parameter vectors (``genotypes'') is disturbed ("mutated") at each iteration (``generation,'' and their objective function value (``fitness'') is assessed. The population for the next generation is created by recombining the highest-scoring parameter vectors, and this process is repeated until the goal is completely optimized. ES can be used to search for a problem's feasible solution space and then discover the best possible solution, which is uncertainty optimization in the optimization issue. 
\par ES has successfully solved parameter optimization problems in adversarial attack generation research. The Authors in \cite{Andrei} used the Evolution search approach to build an untargeted black-box adversarial attack to minimize $ L_{0} $ adversarial perturbations in image setup. The Authors in \cite{Leonardo} compares the development of black-box adversarial attacks for neural network image classification applications using three well-known Evolution techniques. The covariance matrix adaptation evolution strategy (CMA-ES) outperformed the other two strategies in discovering adversarial attacks with tiny perturbations. Our approach is to employ CMA-ES as a learning agent to find the best perturbation parameters that cause the Ransomware classifier to misclassify in binary data scenarios.

\subsection{Generative Adversarial Network}
\label{gan} 
\par Generative Adversarial Network (GAN) is a form of deep learning system that trains two models simultaneously: a generator and a discriminator. The generator's goal is to capture the distribution of specific target data. The discriminator aids in the training of the generator by evaluating how closely the data created by the generator mirrors the original input vector, thus assisting the generator in learning the distribution underlying the genuine input data. GANs are commonly used in the field of image or video production, where providing a sufficient quantity of input images allows the GAN to generate a series of images that resemble but are distinct from the input. In other words, it learns from the intrinsic characteristics of the input, making it a versatile and powerful tool. Malware creators have also employed GANs to generate adversarial attacks. MalGAN\cite{Weiwei} and Pesidious\cite{pesidious} are two such examples. Inspired by these efforts, this study uses a GAN to produce an adversarial feature vector that manipulates a ransomware executable (.exe) file to appear benign. 

\begin{figure*}[htbp]
	\centering
	\includegraphics[width=0.8\textwidth,keepaspectratio]{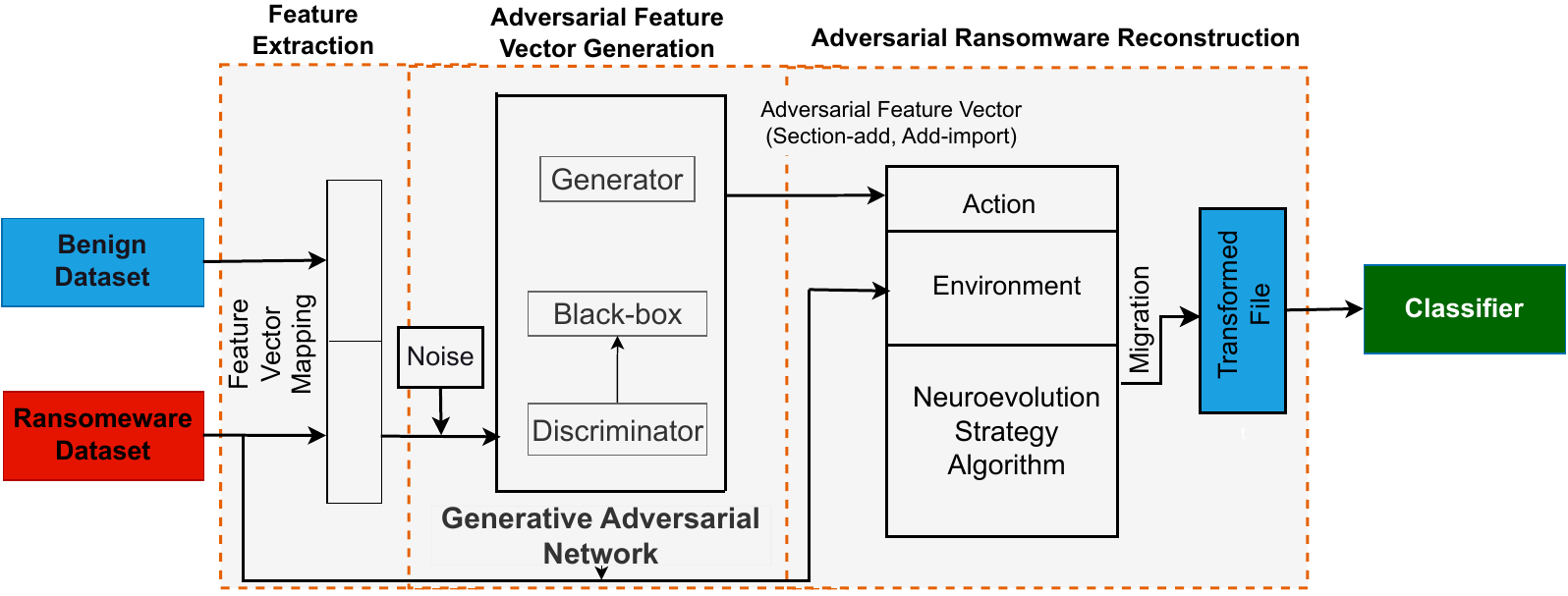}
	\caption{Overview of EGAN, an Evolution GAN adversarial Ransomware Examples Generator. 
 }
	\label{fig:framW}
\end{figure*}

\section{Methodology}
\label{REGAN} 
\par Our method employs an Evolution Strategy (ES) and a Generative Adversarial Network (GAN) to optimize actions with the aim of maximizing evasive potency. Figure~\ref{fig:framW} provides an overview of our framework, dubbed EGAN, which comprises three steps: (a) feature extraction, (b) generation of vectors with positive characteristics, and (c) generation of positive malware. Given a benign and ransomware dataset, we extract features which are then passed to a GAN to generate individual feature mappings for each ransomware and benign PE file. These features include sections and imports. Using the available feature vectors, the GAN engine combines the feature vectors with a noise vector to form an adversarial feature vector for the sections and imports. The Evolution Algorithm agent, alongside the environment, learns from these inputs based on the action selected by the agent. The final output from the agent mutates the malware sample and is sent to a black-box classifier for scoring.

\subsection{\textbf{Feature extraction}} 
\par This study focuses on ransomware that uses the Portable Executable (PE) file format in the Windows operating system family (specifically, Windows PE malware). While ransomware affects many operating systems and various file formats, we chose to focus on Windows for two reasons: (1) according to a 2021 Kaspersky Lab analysis, Windows is the most widely used operating system among end users, and ransomware in the PE file format is one of the most well-known and widely researched threats today. (2) Despite ransomware's diverse file formats, we contend that the knowledge and methodology behind Windows PE ransomware can easily be adapted and applied to other types of ransomware built on different file formats in various operating systems \cite{Darabian}, such as Linux or Android ransomware. The Portable Executable (PE) format is employed by both 32-bit and 64-bit Windows operating systems for executables, object code, DLLs, and other file types. The PE file consists of numerous components, but this study's feature extraction process focuses on the section tables in the headers and the import functions in the section compartment. The section names and import function features are extracted from both ransomware and benign samples. We utilize the same feature extraction process described in \cite{pesidious,Zhiyang,Anderson,Anderson2017EvadingML,Weiwei}. We employ the hashing trick concept to collapse finite features into a 518-dimensional vector, which is then normalized to a value between -0.5 and 0.5. These section names and import features from a PE file can also be used to train classifiers against ransomware, as they provide a holistic view of these attack samples.

\subsection{\textbf{Adversarial feature vector generation}} 
\par The GAN model in EGAN comprises a black-box detector and two neural networks: a Generator and a Discriminator, which are trained in an adversarial situation to capture the distribution of the input feature set.
\par The generator takes in the feature vectors and randomly generated noise (i.e., in the form of 0s and 1s) as input. It then alters this input to generate an adversarial feature vector. The GAN Generator produces new adversarial feature vectors of the same size as the input vector, thus acting as a synthetic data generator. ES uses these vectors as input when the agent selects the appropriate action.
\par Conversely, the Discriminator learns the approximation of the decision function of the black-box detector. This differential function is then provided to the generator to construct a better gradient for learning. 
\par The black-box detector's goal is to determine whether the vector of adversarially generated characteristics is malicious or benign. This component is distinct from the other elements of the generative network and is never retrained. The black-box detector is trained using a Random Forest model with 100 estimators. Readers can refer to \cite{Weiwei,pesidious,Anderson2017EvadingML} for a detailed description of the GAN model.

\begin{table}[htbp]
\caption{Actions used in EGAN}
\centering 
	\scalebox{0.85}{
 \begin{tabular}{|c|} 
		\hline    
		ACTION\_TABLE  \\ 
		\hline 
		'section\_rename': 'section\_rename' \\
		'section\_add' : 'section\_add'  \\
		'add\_imports' : 'add\_imports' \\  
		'append\_benign\_binary\_overlay' :'append\_benign\_binary\_overlay'\\ 
		\hline
	\end{tabular}
 }
	\label{tab:actions}
\end{table}

\subsection{\textbf{Adversarial Ransomware generation}} 
\par The Adversarial Ransomware attack is considered a stochastic optimization problem involving an environment and an agent acting within this environment. Specifically, in EGAN, the Covariance Matrix Adaptation Evolution Strategy (CMA-ES) is employed as our agent, which is controlled by actions, and each action manipulates a set of features.
\par The implementation of the CMA-ES algorithm in EGAN is based on the work presented in \cite{Salimans}. It utilizes parallelization, z-normalization fitness shaping, and a two-layer neural network as a policy network to generate the most optimal sequence of mutations to apply on a Ransomware sample. 
\par During the CMA-ES algorithm training and observations from the literature, it was discovered that actions such as renaming/adding sections, adding imports, and appending benign binary overlays reduced the true positive rate of the Ransomware while preserving the semantics of the malware file. As a result, only these four actions were considered in EGAN to train the agent. A complete list of these actions is provided in Table~\ref{tab:actions}. The Adversarial Ransomware generation process is as follows: 
\begin{itemize}
    \item If 'section\_add', 'add\_imports', or 'section\_rename' is selected as the action, the agent utilizes the adversarial feature vector created by the GAN for the imports and sections as input. Due to the large feature space of the sections and imports in a PE file, an agent learning directly from this large pool would result in an enormous amount of manipulations and training time. Hence, the presence of GAN in EGAN restricts the agent learning process.
    \item If 'appending\_benign\_binary\_overlays' is randomly selected by the agent, these contents are sourced directly from benign files, not from the GAN. GAN, by nature, cannot be used to generate contents not inherently present in files, such as appending one file to another. We believe that adding content taken directly from benign programs would deceive the classifier into computing the probability of being Ransomware.
\end{itemize}
\par The agent mutates the Ransomware sample and calculates a reward based on the actions. The mutation is done in conjunction with an environment. EGAN utilizes the OpenAI Gym-malware \textit{``malware-score-0v}'' environment \cite{Anderson}. After every interval, a batch of experiences sampled using priorities is employed to calculate the loss, which is then backpropagated to a deep neural network (policy network) to update the weights.
\par The agent is tested after a certain number of episodes to check the success rate. If it exceeds a threshold, the training is halted, and the model is saved for use in mutating Ransomware. The CMA-ES model is trained to select the best parameters along with other tools to generate new samples that can fool a black-box classifier. The LIEF tool \cite{LIEF} is used to apply these actions and make modifications to the Ransomware file.

\section{Experiments}
\label{exp}
\par Our experiment runs in a black-box environment on a Kali GNU/Linux Rolling machine, version 2020.4, with an Intel Core i5 processor and 8GB of RAM. All EGAN scripts are written in Python 3.7, with the exception of the binary reconstruction script, which is written in the C++ library. 
\par The Ransomware samples used for training and testing the solution were sourced from GitHub and various other online platforms. We used a total of 150 Ransomware samples, and approximately 2500 benign samples for this experimentation. Due to significantly low response times from the VirusTotal API, Kaspersky Threat Intelligence Portal, and Cuckoo sandbox, we limited the number of generated mutated examples for evaluation to popular Ransomware examples listed online.

\begin{figure*}[htbp]
	\centering
	\includegraphics[width=\textwidth,height=\textheight]{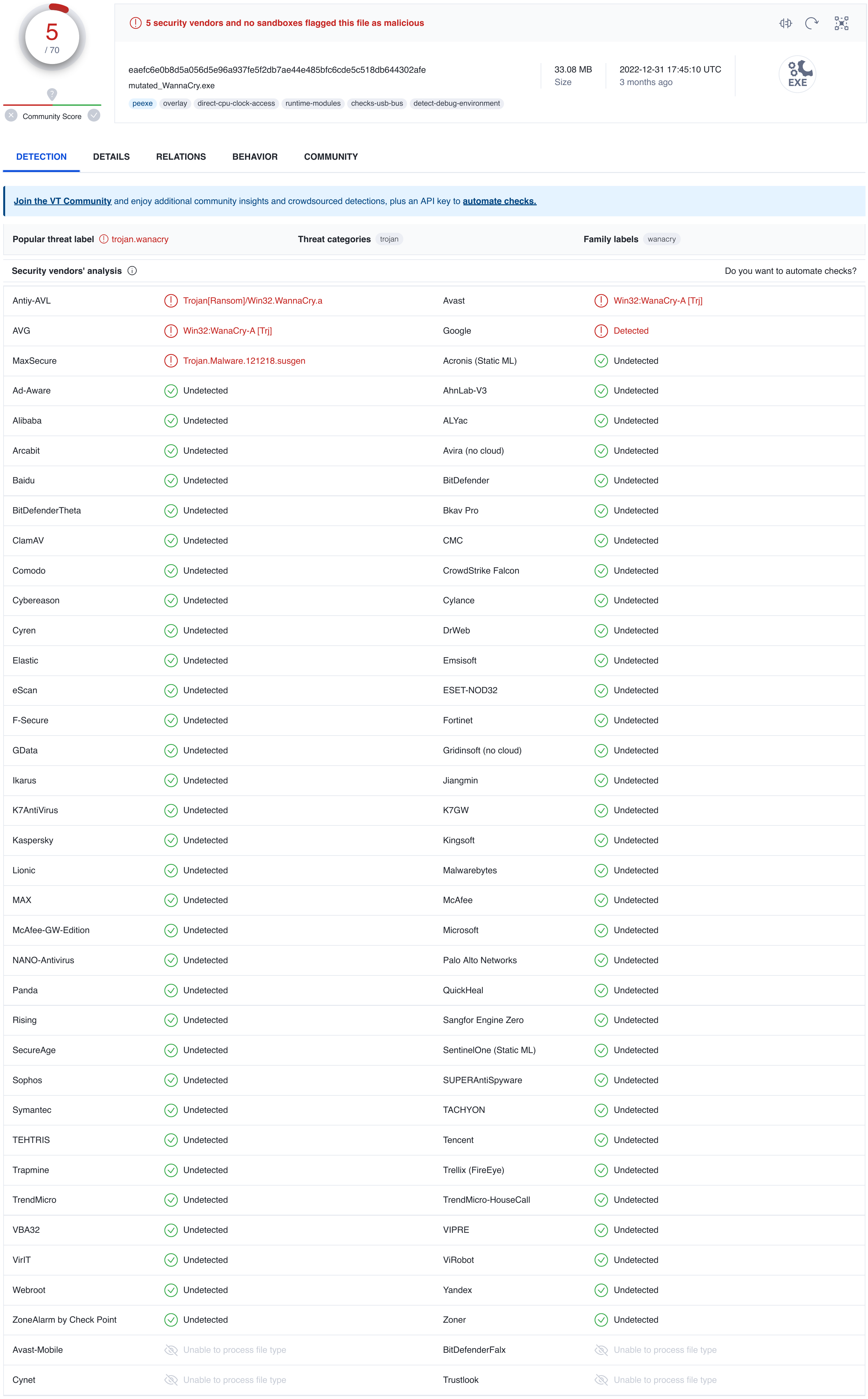}
	\caption{Screenshot of VirusTotal scanned results showing EGAN evasive against popular AI-powered AV. \url{https://t.ly/gbaC}}
	\label{fig:VT}
\end{figure*}

\begin{figure*}[htbp]
	\centering
	\begin{subfigure}[b]{0.48\linewidth}
		\includegraphics[width=\linewidth]{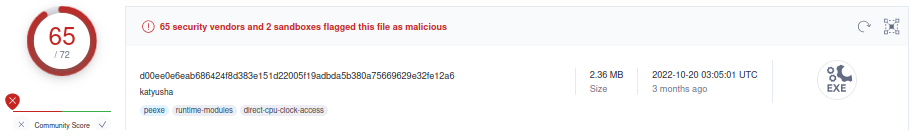}
		\caption{$Katyusha$}
	\end{subfigure}
	\begin{subfigure}[b]{0.48\linewidth}
		\includegraphics[width=\linewidth]{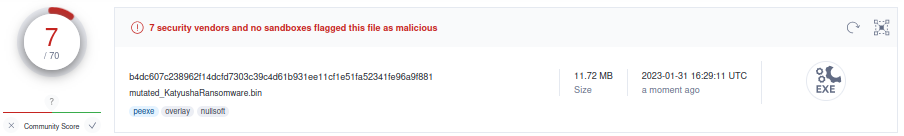}
		\caption{$Ad\_Katyusha$}
	\end{subfigure}

   \begin{subfigure}[b]{0.48\linewidth}
		\includegraphics[width=\linewidth]{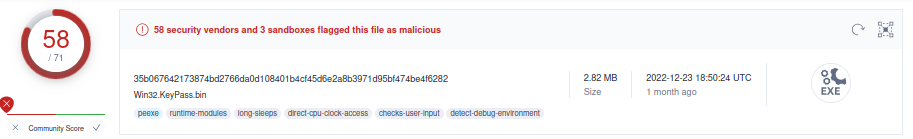}
		\caption{$Keypassare$}
	\end{subfigure}
	\begin{subfigure}[b]{0.48\linewidth}
		\includegraphics[width=\linewidth]{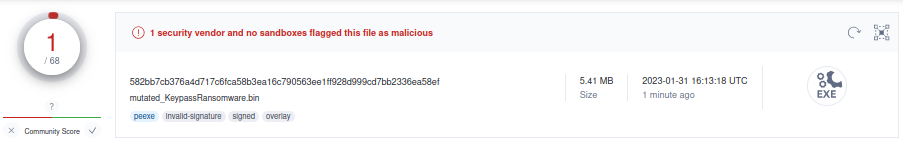}
		\caption{$Ad\_Keypass$}
	\end{subfigure}
	
	\begin{subfigure}[b]{0.492\linewidth}
		\includegraphics[width=\linewidth]{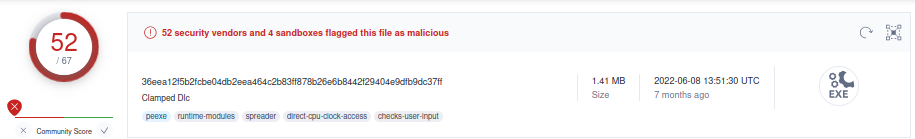}
		\caption{$ Kryptik$}
	\end{subfigure}
	\begin{subfigure}[b]{0.48\linewidth}
		\includegraphics[width=\linewidth]{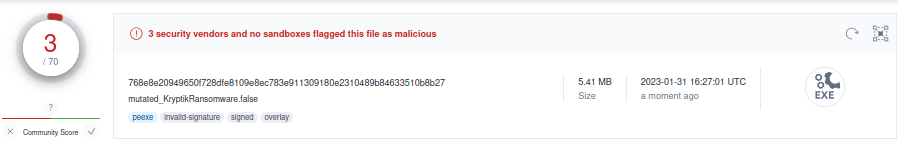}
		\caption{$Ad\_Kryptik$}
	\end{subfigure}
	
	\begin{subfigure}[b]{0.48\linewidth}
		\includegraphics[width=\linewidth]
        {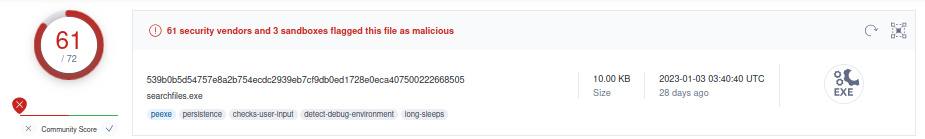}
		\caption{$ LockCrypt2.0 $}
	\end{subfigure}
	\begin{subfigure}[b]{0.48\linewidth}
		\includegraphics[width=\linewidth]
        {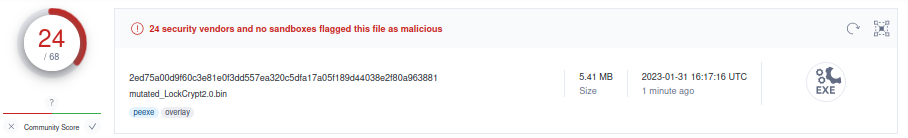}
		\caption{$Ad\_LockCrypt2.0 $}
	\end{subfigure}
	
	\begin{subfigure}[b]{0.48\linewidth}
		\includegraphics[width=\linewidth]{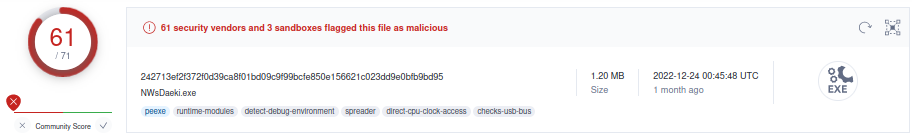}
		\caption{$ Marix$}
	\end{subfigure}
	\begin{subfigure}[b]{0.48\linewidth}
		\includegraphics[width=\linewidth]{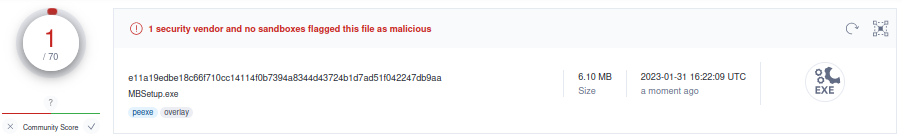}
		\caption{$Ad\_Marix$}
	\end{subfigure}	
	
	\begin{subfigure}[b]{0.48\linewidth}
		\includegraphics[width=\linewidth]
        {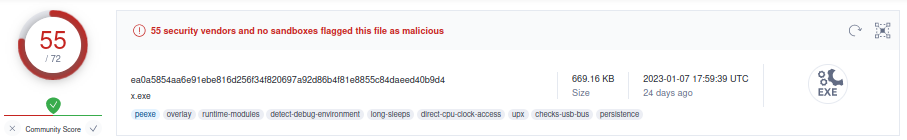}
		\caption{$ Moon $}
	\end{subfigure}
	\begin{subfigure}[b]{0.48\linewidth}
		\includegraphics[width=\linewidth]
        {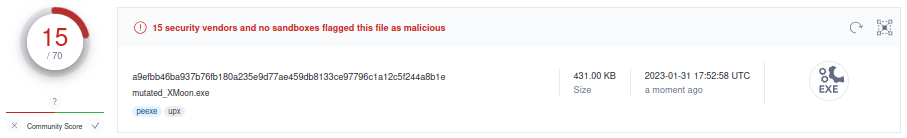}
		\caption{$Ad\_Moon $}
	\end{subfigure}
	
	\begin{subfigure}[b]{0.48\linewidth}
		\includegraphics[width=\linewidth]
        {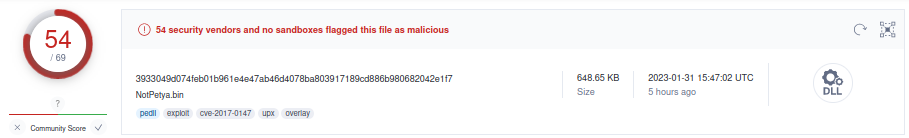}
		\caption{$ Pataya $}
	\end{subfigure}
	\begin{subfigure}[b]{0.48\linewidth}
		\includegraphics[width=\linewidth]
        {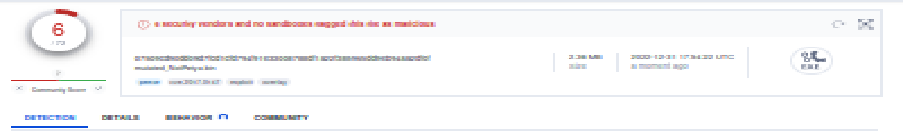}
		\caption{$Ad\_Pataya $}
	\end{subfigure}
	
	\begin{subfigure}[b]{0.48\linewidth}
		\includegraphics[width=\linewidth]
        {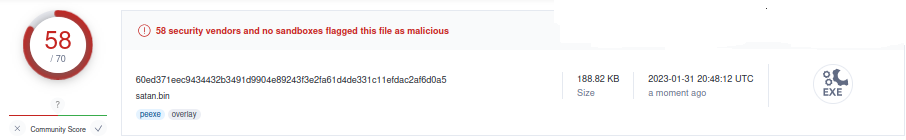}
		\caption{$ Satan $}
	\end{subfigure}
	\begin{subfigure}[b]{0.48\linewidth}
		\includegraphics[width=\linewidth]
        {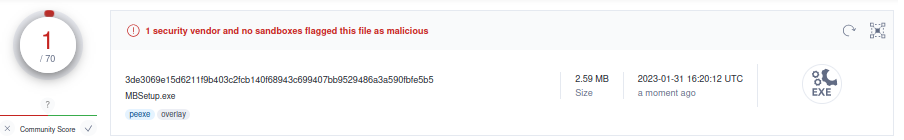}
		\caption{$Ad\_Satan $}
	\end{subfigure}
	
	\begin{subfigure}[b]{0.48\linewidth}
		\includegraphics[width=\linewidth]
        {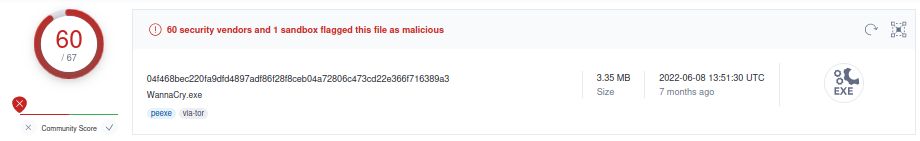}
		\caption{$ WannaCry $}
	\end{subfigure}
	\begin{subfigure}[b]{0.48\linewidth}
		\includegraphics[width=\linewidth]
        {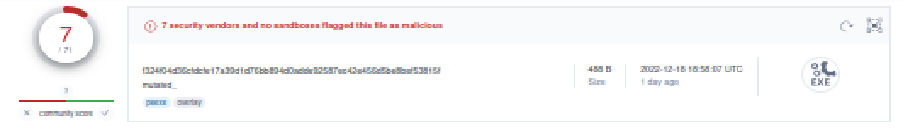}
		\caption{$Ad\_WannaCry $}
	\end{subfigure}

	\caption{VirusTotal scanned results for popular Ransomware samples and their adversarial counterparts.}
	\label{fig:3}
\end{figure*}

\begin{figure*}[htbp]
	\centering
	\includegraphics[width=7.11in]{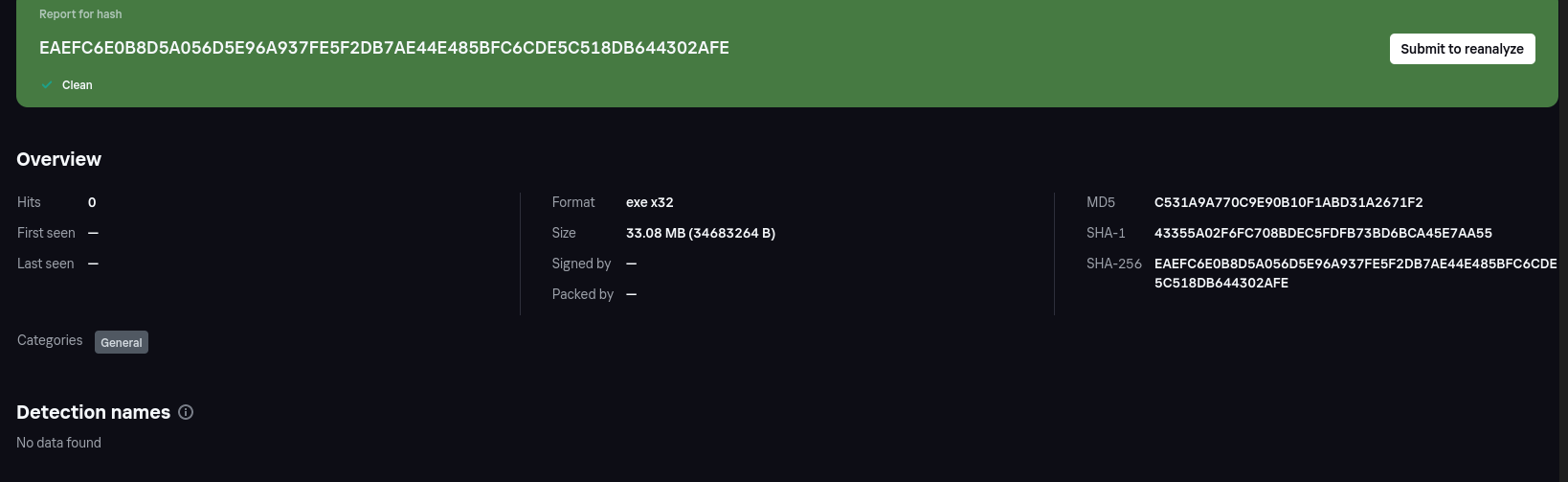}
    \captionsetup{justification=centering}
	\caption{Kaspersky detection results for transformed Ransomware. \\
    The Kaspersky Threat Intelligence portal found no data on this file. \url{https://t.ly/O8KA}}
	\label{fig:kaspasky}
\end{figure*}

\begin{figure*}[htbp]
	\centering
	\includegraphics[width=7.11in]{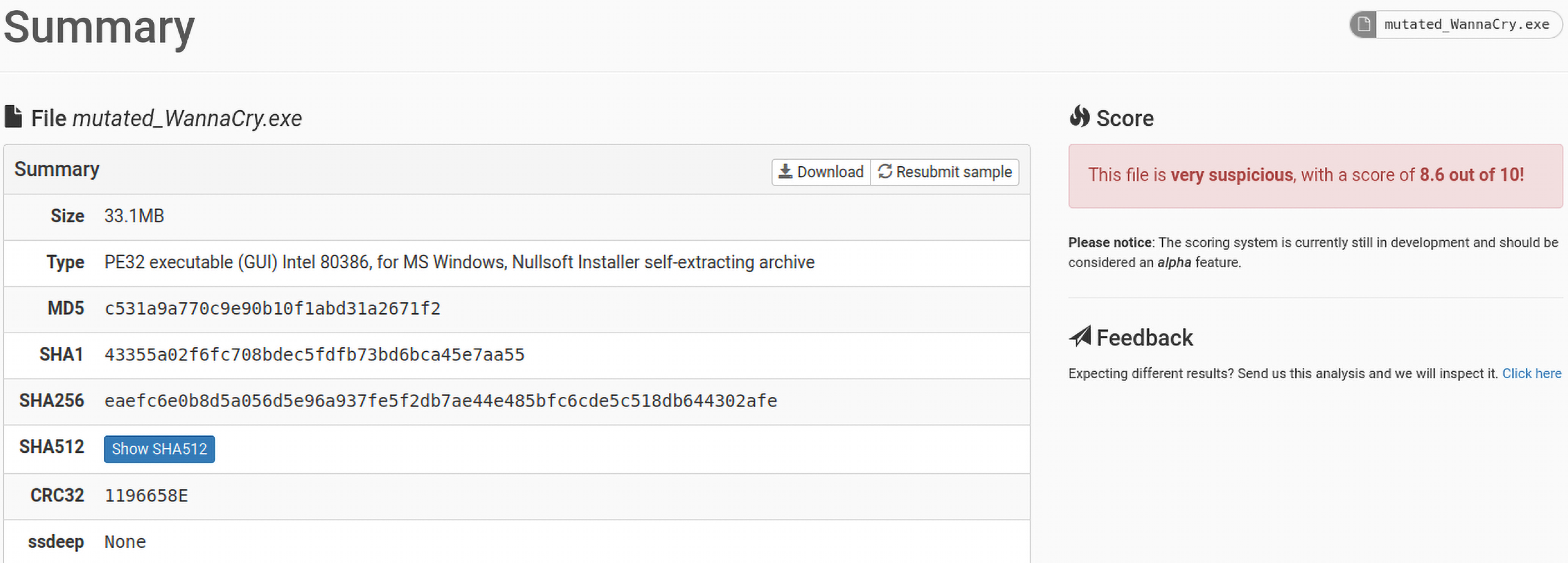}
    \captionsetup{justification=centering}
	\caption{Screenshot of Cuckoo sandbox report.}
	\label{fig:cockoo}
\end{figure*}

\begin{figure*}[htbp]
	\centering
	\includegraphics[width=7.11in]{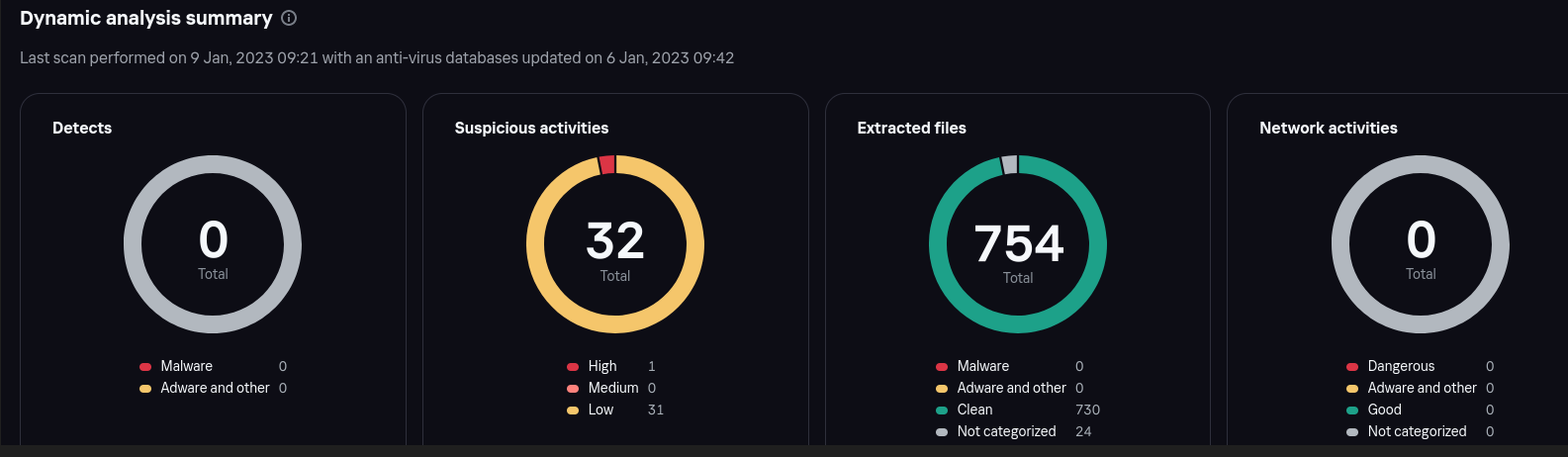}
    \captionsetup{justification=centering}
	\caption{Screenshot of Kaspersky dynamic analysis summary. \\
    No sandbox detected the mutated Ransomware file.\url{https://t.ly/O8KA}}
	\label{fig:kdynamic}
\end{figure*}

\par The GAN structure adheres to the implementation settings presented in \cite{Weiwei,pesidious}. The policy network for the CMA-ES agent comprises two linear layers {256, 64}, each followed by a rectified linear unit function. We limit the training of the CMA-ES agent to the four manipulations specified in Table~\ref{tab:actions}. The CMA-ES agent receives a reward for each manipulation based on the score of a pre-trained Gradient Boosting model used as a black-box classifier \cite{Anderson2017EvadingML}. This black-box classifier differs from the one implemented in the GAN. The reward is calculated as the difference between the score of the original Ransomware sample and the mutated sample after every action. The episode concludes once the score falls below a threshold of 80
\\This experiment comprises two parts:
\begin{itemize}
	\item We assess the evasive strength of each action and the status of functional preservation for each mutated example resulting from these actions during training.
The results of this experiment are presented in Table~\ref{tab:4}. Mutated examples were uploaded to the Cuckoo sandbox to determine their functionality; if an example executes and generates dynamic features, it is considered functional. The action `appending\_benign\_binary\_overlays' yielded the best performance across all domains.
	
	\item We examine the transferability strength of the adversarial example on other AI-based models. If an adversarial example is transferable, it can evade other anti-Ransomware engines within the same domain after successfully evading the pre-trained Gradient Boosting black-box classifier. We evaluate transferability using popular Ransomware examples (e.g., WannaCry, LockCrpt2.0, Moon, KatyshaRansomware, KeypassRansomware, Pataya, KryptikRansomware, etc.). The rationale for these selections is that these specific Ransomware samples are known to trigger AVs; as a result, if an AV or sandbox detects them, it will immediately flag the file as malicious in both dynamic and static analyses. The EGAN attack is applied to these examples to prevent AVs and sandboxes from detecting them. After the transformation, EGAN reconstructs the file into executable format and it is uploaded to VirusTotal, the Kaspersky Threat Intelligence Portal, and the Cuckoo sandbox, all of which are publicly available on the Internet.  
\end{itemize}

\begin{table}[htbp]
\caption{Evasive rate of each action}
\centering 
\scalebox{0.699}{
 \begin{tabular}{|c|c|c|} 
		\hline    
		Actions &  Functional Examples & Average VirusTotal score\\ 
		\hline 
		section\_rename &   5/5 & 41/70\\
		section\_add  &   4/5 &  32/70\\
		add\_imports  &   4/5 & 39/70\\  
		append\_benign\_binary\_overlay  &  5/5 & 5/70\\ 
		\hline
	\end{tabular}
 }
	\label{tab:4}
\end{table} 

\subsection{Bypassing Static AI-powered commercial antivirus}
\par Several commercial antivirus (AV) systems have adopted Machine Learning/Deep Learning (ML/DL) or Artificial Intelligence (AI) techniques to cope with the relentless proliferation of new Ransomware strains. This is due to these techniques' ability to generalize and recognize previously unseen malicious Ransomware strains \cite{George,Hyrum,Edward,Joshua,Konrad}. However, we demonstrate that transformations on Ransomware samples that infiltrate the black-box classifier can also compromise many ML/AI-based detectors. We tested our hypothesis on over sixteen (16) commercial ML-powered detectors listed on VirusTotal, relying on the responses retrieved from the platform. Figure~\ref{fig:VT} illustrates the results of this experiment. From the figure, it can be inferred that the adversarial Ransomware was effective against all scanners except Max-Secure and Avast. The sixteen (16) other scanners were susceptible to the transformations we used in our investigation, despite the attack not being specifically designed against them.

\subsection{Bypass other Static Commercial Scanners}
\label{other}
\par We utilize each antivirus detection output to evaluate the effectiveness of EGAN. We argue that a binary is considered benign if VirusTotal and other similar engines deem it as such, and if the dynamic analysis from a sandbox does not generate an alert. Consequently, the ransomware can evade detection and be used by an individual user. As illustrated on the left side of Figure~\ref{fig:3}, multiple antivirus engines identify Ransomware executables as malicious in the absence of an attack. The final executable, following the transformation, achieves lower detection from VirusTotal; see the right side of Figure~\ref{fig:3}. These results are further cross-validated with the Kaspersky Threat Intelligence Portal's multi-engine scanner, as shown in Figure~\ref{fig:kaspasky}. Based on the aforementioned results, it is clear that by making certain "static" modifications to the Windows PE executable, we can circumvent the static analyses of many AVs.   

\subsection{Dynamic Analysis with Sandboxes}
\par The dynamic analysis is intended solely to bypass the security checks that are run when the binary is executed in a Cuckoo sandbox and Kaspersky endpoint environment. The goal of the ``$ append\_benign\_binary\_overlay $'' actions on the adversarial Ransomware is to reduce the likelihood of reaching the detection threshold. In particular, this action embeds benign executables into the Ransomware loader. The loader subsequently drops the benign executable and decoded data, generating a new process to run the benign executable. Once the benign executable has run, the loader reverses the transformation on the Ransomware and initiates it. This process confounds the sandbox because many positive indicators are derived from the benign executable; all connections and activities conducted by it are already whitelisted, leading the sandbox to label the entire Ransomware loader as a benign executable. Upon successful implementation of the actions mentioned above, the Ransomware is launched. Alarmingly, our evasion methods were effective, as shown by the low scores achieved in Kaspersky Threat Intelligence Portal sandboxes (see Figure~\ref{fig:kdynamic}). However, the performance was less satisfactory in the Cuckoo sandbox (see Figure~\ref{fig:cockoo}), with a detection score of 8.6 out of 10.
\par The Cuckoo sandbox is also used to ensure that the transformed Ransomware maintains its malicious behavior after the actions have been applied. The Cuckoo sandbox collects sample behaviors from the Ransomware and translates them into comprehensible descriptive signatures. Each signature is a text string that encapsulates a particular sample behavior. We compare the actions of the modified Ransomware to those of the original. We classify a Ransomware variant as evasive if it behaves identically to the original. The behavioral similarity between two payloads is defined as both samples sharing the same behavioral functions. If this is not the case, we infer that the transformation has altered the behaviors of the original Ransomware. In Figure~\ref{fig:cockoo}, with a score of 8.6 out of 10, there is no need to compare similarities, since the transformed Ransomware has maintained its original functionality; if it had not, the sandbox would not have achieved such a high detection rate.

\section{Discussion and Conclusion}
\label{concl} 
\par Identifying an efficient method to generate Adversarial Ransomware for Adversarial Training without compromising its functionality remains a challenging task. However, this study presents a compelling argument for using EGAN (Evolution Strategy with GAN) to generate adversarial Ransomware. The adversarial Ransomware samples were evaluated against both static and dynamic Ransomware classifiers, with the transformations applied to the Ransomware achieving a notable evasion rate.
\par The findings presented in the previous sections highlight the concerning nature of adversarial Ransomware. Specifically, the aforementioned static analysis confirms that Ransomware classifiers are ineffective at detecting these types of attacks, a fact contrary to common customer beliefs. Furthermore, these static classifiers are typically designed to address specific problem sets, operating under the assumption that their training and test data originate from the same statistical distribution. Unfortunately, in high-stakes applications, this assumption is frequently violated in critical ways, as the complete transformation contradicts these statistical assumptions, resulting in the high evasion rates recorded by these classifiers. During testing, well-known antivirus programs on VirusTotal and Sandboxes failed to detect adversarial Ransomware when it was uploaded and executed. However, the generated Ransomware samples can also be used to train classifiers and detectors (also known as Adversarial Training) because both dynamic and static features can be extracted from these Ransomware strains. 
\par In future research, we plan to investigate other actions and additional structures of PE file exploitation that can evade dynamic analysis. Our experimentation shows that the four actions currently employed lack the robustness needed to evade dynamic analysis from the Cuckoo sandbox. Another significant limitation was the response time from commercial scanners, which restricted the amount of data samples used in our analysis. 

\section*{Acknowledgment}
This paper was presented at the 48th IEEE Conference on Local Computer Networks (LCN) and was published in the conference proceedings. The final authenticated version is available online at: \url{https://doi.org/10.1109/LCN58197.2023.10223320}

\bibliographystyle{IEEEtran} 
\bibliography{reference} 

\end{document}